\documentclass[10pt, a4paper]{article}

\usepackage[final]{lrec2026} 

\usepackage{mdframed}
\usepackage{colortbl} 
\usepackage{placeins}
\usepackage{booktabs}
\usepackage{multirow}
\usepackage{tcolorbox}
\usepackage{enumitem}
\usepackage{subcaption}
\usepackage{amsmath}
\usepackage{amsthm}
\usepackage{amsfonts}
\usepackage{xspace}
\usepackage{longtable}
\usepackage{array}
\usepackage{tabularx}
\usepackage{siunitx}
\usepackage{bm}
\usepackage{graphicx}
\usepackage{pifont}

\newtcolorbox{mybox}
{colback=red!5!white,colframe=red!75!black}

\title{PhonemeDF: A Synthetic Speech Dataset for Audio Deepfake Detection and Naturalness Evaluation}

\name{\shortstack[c]{
Vamshi Nallaguntla$^{1,*}$, Aishwarya Fursule$^{1}$, Shruti Kshirsagar$^{1}$,\\
Anderson R. Avila$^{2,3}$}}

\address{
$^1$ Wichita State University, Wichita, Kansas, USA\\
$^2$ Institut national de la recherche scientifique (INRS-EMT), Université du Québec, Canada\\ 
$^3$ INRS-UQO Mixed Research Unit on Cybersecurity, Gatineau, Canada\\
\{vxnallaguntla@shockers., axfursule@shockers., shruti.kshirsagar@\}wichita.edu,\\ Anderson.Avila@inrs.ca}

\abstract{The growing sophistication of speech generated by Artificial Intelligence (AI) has introduced new challenges in audio deepfake detection. Text-to-speech (TTS) and voice conversion (VC) technologies can create highly convincing synthetic speech with naturalness and intelligibility. This poses serious threats to voice biometric security and to systems designed to combat the spread of spoken misinformation, where synthetic voices may be used to disseminate false or malicious content. While interest in AI-generated speech has increased, resources for evaluating naturalness at the phoneme level remain limited. In this work, we address this gap by presenting the Phoneme-Level DeepFake dataset (PhonemeDF), comprising parallel real and synthetic speech segmented at the phoneme level. Real speech samples are derived from a subset of LibriSpeech, while synthetic samples are generated using four TTS and three VC systems. For each system, phoneme-aligned TextGrid files are obtained using the Montreal Forced Aligner (MFA). We compute the Kullback--Leibler divergence (KLD) between real and synthetic phoneme distributions to quantify fidelity and establish a ranking based on similarity to natural speech. Our findings show a clear correlation between the KLD of real and synthetic phoneme distributions and the performance of classifiers trained to distinguish them, suggesting that KLD can serve as an indicator of the most discriminative phonemes for deepfake detection.
 \\ \newline  }

\begin{document}

\maketitleabstract

\begingroup
\renewcommand\thefootnote{}
\footnotetext[0]{* Corresponding author.}
\endgroup

\section{Introduction}

Early synthesizers generated speech signals that sounded unnatural and robotic, making them easily recognizable \cite{nusbaum1997measuring}. This remained the case until recent advancements in generative artificial intelligence \cite{ren2019fastspeech, prenger2019waveglow}, which led to improvements in the quality of synthetic speech. In fact, the growing sophistication of Text-to-Speech (TTS) and Voice Conversion (VC) systems can benefit society in multiple ways. Besides their impact on customer experience, high-quality synthetic speech can help expand accessibility to many people \cite{naayini2025ai}. TTS, specifically, enables individuals with visual impairments or reading disabilities to access written content, such as e-books, websites, and documents. VC, on the other hand, improves personalization and communication by adapting voices for specific characters, accents, or age groups in the entertainment domain.

Despite its benefits, the generation of high-quality synthetic speech has introduced new challenges in spoofing detection. Deepfakes, in particular, pose significant challenges not only for voice biometric security applications but also for systems designed to combat the spread of spoken misinformation, where false content is deliberately generated using the voice of a targeted individual for malicious purposes \cite{yamagishi2021asvspoof}. In such a context, the ASVspoof challenge has been the main force providing the research community with guidelines and resources to promote the robustness of ASV systems \cite{wu2015asvspoof}. More recently, the organizers, realizing the rapid development of deep learning techniques used for voice conversion and speech synthesis, added the audio deepfake detection task to the ASVspoof 2021 challenge. In the following year, the Audio Deep Synthesis Detection (ADD) challenge was created by another group to cover attacks performed in more realistic scenarios, such as those involving background noise, fake clips in real speech signals, and new speech synthesis and voice conversion algorithms \cite{yi2022add}.  

The main objective of the participants in such challenges is to develop solutions that can surpass the performance of baseline systems, creating new benchmarks for the task at hand. The focus is placed on developing new approaches for front-end and back-end systems. Besides the numerous contributions throughout the years, such approaches often lack a deep investigation of the phenomenon. In this work, we aim to fill this gap by studying the main acoustic differences between real and synthetic speech. We focus our analysis at the phoneme level to minimize the effects of linguistic variability. Our main objective is to investigate speech parameters associated with naturalness (e.g., speech rate, intensity range, articulation rate, average syllable duration) at the phoneme level and to compare the presence of these parameters in real and synthetic speech. Once these parameters are well understood, a model with an internal extractor of these parameters can be developed to compute a new naturalness score. In this work, we propose a comprehensive analysis framework to assess speech naturalness through phoneme-level alignment. We construct a parallel dataset comprising real speech from the LibriSpeech corpus and synthetic speech generated using four text-to-speech (TTS) models: MeloTTS~\citep{melotts2024}, XTTS v2~\citep{xttsv2_2024}, Chatterbox TTS~\citep{Chatterbox2024}, and VITS TTS~\citep{VITS2021}, and three voice conversion (VC) models: Chatterbox VC~\citep{Chatterbox2024}, FreeVC~\citep{FreeVC-23}, and StarGAN VC~\citep{StarGANv2VC-21}. Our findings highlight specific phonemes and acoustic patterns that contribute most to perceived unnaturalness. Aligned with that, recent phoneme-aware studies \citep{suthokumar2019phoneme, dhamyal2021using, sivaraman2025voiced} highlight that certain phoneme categories exhibit stronger separability between real and fake speech. Yet, existing datasets offer only utterance- or frame-level labels, lacking aligned phoneme boundaries or timestamps, which makes phoneme-based analysis and reproducibility challenging. To address this gap, we present a new PhonemeDF dataset. In particular, in this paper, the following contributions are made:
\begin{enumerate}
    \item We introduce PhonemeDF, a new dataset containing synthetic audio and time-aligned phoneme boundaries extracted from TextGrids for both real and synthetic recordings.
    \item We develop a phoneme-level detection framework that evaluates discriminability across phonemes using Kullback--Leibler divergence (KLD), and ML-based classifiers such as Logistic Regression (LR) and Support Vector Machine (SVM).
    \item Lastly, we conduct a comparative study across handcrafted features and self-supervised learning (SSL) embeddings to assess their robustness and generalization in phoneme-level deepfake detection.
\end{enumerate}

\section{Related Work}

\subsection{Naturalness Assessment Based on Phonemes}
The concept of naturalness is not clearly defined in the literature \cite{nussbaum2025understanding}. Our ability to perceive robotic characteristics in synthetic speech can be quite subjective \cite{nusbaum1997measuring}, and has decreased in recent years, given the advances in generative AI, which is now capable of generating high-quality speech. Studies addressing speech naturalness date back to the early nineties. In \cite{nusbaum1997measuring}, for instance, the authors proposed a new methodology for measuring the naturalness of specific aspects of synthesized speech, independent of its intelligibility. While naturalness is a multidimensional and subjective quality of speech, this methodology enables the assessment of the distinct contributions of prosodic, segmental, and source characteristics within an utterance. They conducted experiments showing that glottal source features and prosodic cues (like pitch and rhythm) help listeners distinguish real from synthetic speech. This provides a foundation for more focused evaluations of speech quality. Although a handful of studies have explored speech naturalness \cite{dall2014rating, sellam2023squid, vojtech2019effects}, to the best of our knowledge, most of them are not focused on phoneme analysis, which is addressed in our study.

\subsection{Deepfake detection datasets}

While large-scale deepfake detection datasets such as ASVspoof \citep{todisco2019asvspoof} and MLAAD \citep{li2024audio} provide extensive collections of real and synthetic speech, they lack phoneme-level annotations. Recent efforts have begun to address this limitation. Baser et al. \citep{baser2025phonemefake} introduced PhonemeFake, a dataset focused on segmental deepfake manipulations, and proposed an adaptive bilevel detection model that achieved lower equal error rates (EERs) with 90\% faster inference compared to prior baselines. Temmar et al. \citep{temmar2025phonetic} created a phoneme-annotated dataset. They developed a phoneme-to-word HuBERT-based framework to classify real vs. synthetic speech. Their analysis showed that diphthongs and fricatives exhibit the strongest deviations and provide interpretable phoneme-level cues for detection. Yang et al. \citep{yang2025forensic} compared phonetic features with global audio-level representations to evaluate which feature types best discriminate genuine from synthetic speech.

\begin{figure*}
\centering
\includegraphics[width=0.8\textwidth]{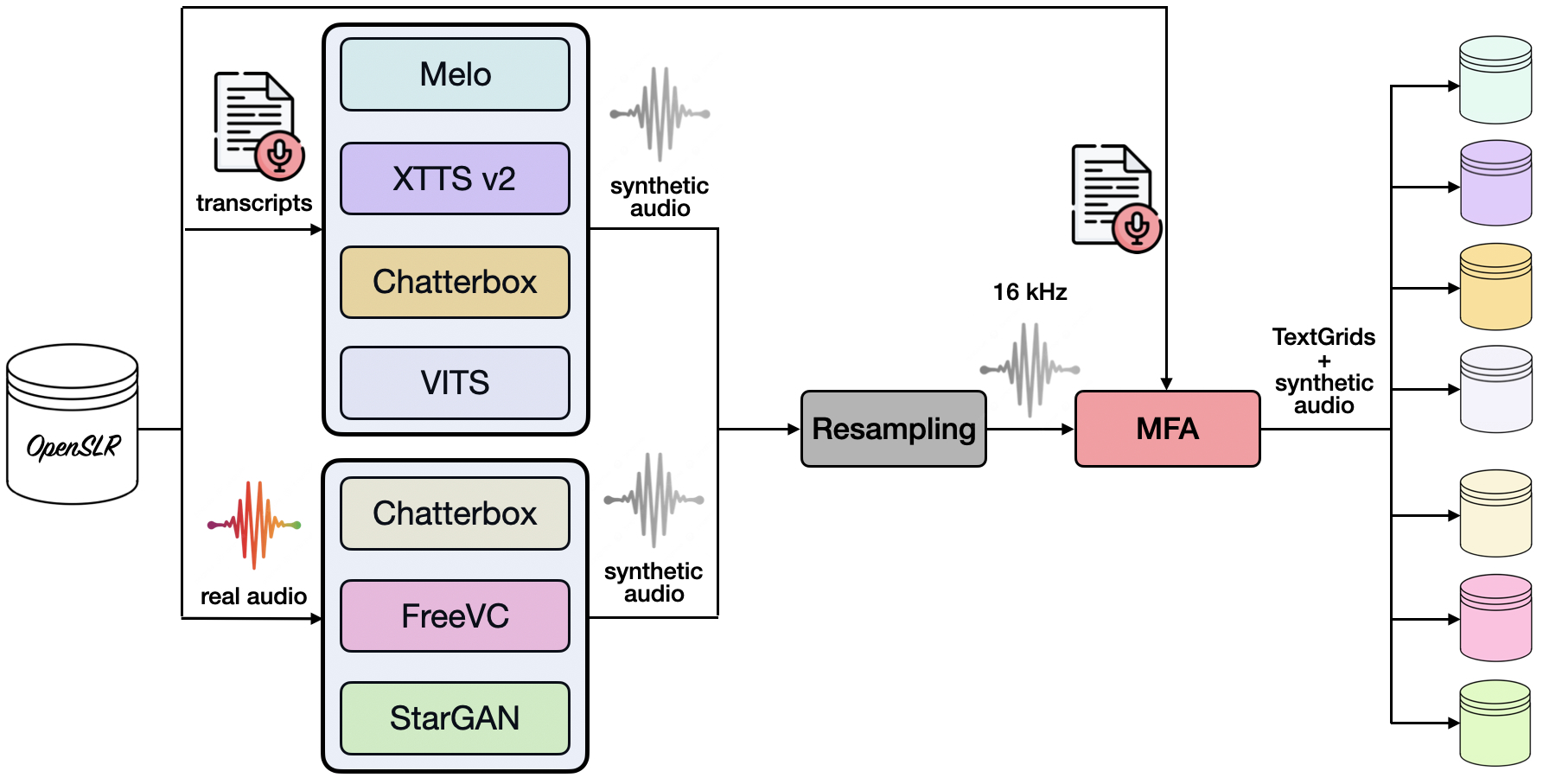}
\caption{Overview of the dataset creation pipeline.}
\label{fig:pipeline1}
\end{figure*}

Beyond dataset creation, several studies have explored phoneme-driven detection strategies. For example, Salvi et al. \citep{salvi2025phoneme} investigated speaker-specific phoneme profiles and evaluated test audio against these profiles to accurately identify synthetic artifacts. For replay attack detection, Suthokumar et al. \citep{suthokumar2019phoneme} demonstrated that phoneme-based models improve detection accuracy by capturing phoneme-dependent acoustic patterns. Dhamyal et al. \citep{dhamyal2021using} in turn employed self-attention mechanisms to identify which phoneme-level spectral features most effectively distinguish real from synthetic speech. Sivaraman et al. \citep{sivaraman2025voiced} adapted the AASIST detector \citep{jung2022aasist} by partitioning speech into voiced and unvoiced segments, finding that voiced segments provide stronger discriminative cues. Zhang et al. \citep{zhang2025phoneme} proposed a phoneme-based detector. They utilized adaptive phoneme pooling and graph attention networks to capture inter-phoneme inconsistencies introduced during synthesis.

Research on discriminative acoustic cues has identified specific phonetic categories and features that reveal synthesis artifacts. The work in \citep{temmar2025phonetic, sivaraman2025voiced}, for example, found that some phonemes, such as fricatives, diphthongs, and voiced segments, show more distinct cues from synthetic speech. Furthermore, work in \citep{Zhu2024SLIM} showed that mismatches between linguistic text and speaking style can provide complementary cues for detecting synthetic speech. These results suggest that phoneme-level analysis could enhance detection accuracy and interpretability by providing greater granularity than utterance-level approaches.

\section{PhonemeDF Dataset}
\label{corpus}

The goal of PhonemeDF is to provide phoneme segmentation of parallel real and synthetic speech, enabling the investigation of differences between real and AI-generated signals at the phoneme level. We hypothesize that generative models are optimal at generating certain phonemes and less efficient at producing others. This can be used to leverage the optimization of audio deepfake classifiers. The procedure to construct the dataset is divided into two stages. The first stage consists of generating synthetic utterances from seven distinct synthesizers, followed by the segmentation of the speech into phoneme-level segments. These two stages are detailed in the next two sections and an overview of our data construction is shown in Figure~\ref{fig:pipeline1}.

\subsection{Synthetic Speech Generation}
We adopted four TTS systems, namely MeloTTS~\citep{melotts2024}, XTTS v2~\citep{xttsv2_2024}, Chatterbox TTS~\citep{Chatterbox2024}, VITS TTS~\citep{VITS2021}, and three VC models, referred to as Chatterbox VC~\citep{Chatterbox2024}, FreeVC~\citep{FreeVC-23}, and StarGAN VC~\citep{StarGANv2VC-21} to generate synthetic speech. These speech generation systems were selected based on their open availability, coverage of modern paradigms, and diversity of synthesis mechanisms. LibriSpeech \citep{LibriSpeech} was used as a reference corpus to generate parallel synthetic data. 

\begin{figure*}
\centering
\includegraphics[width=0.8\textwidth]{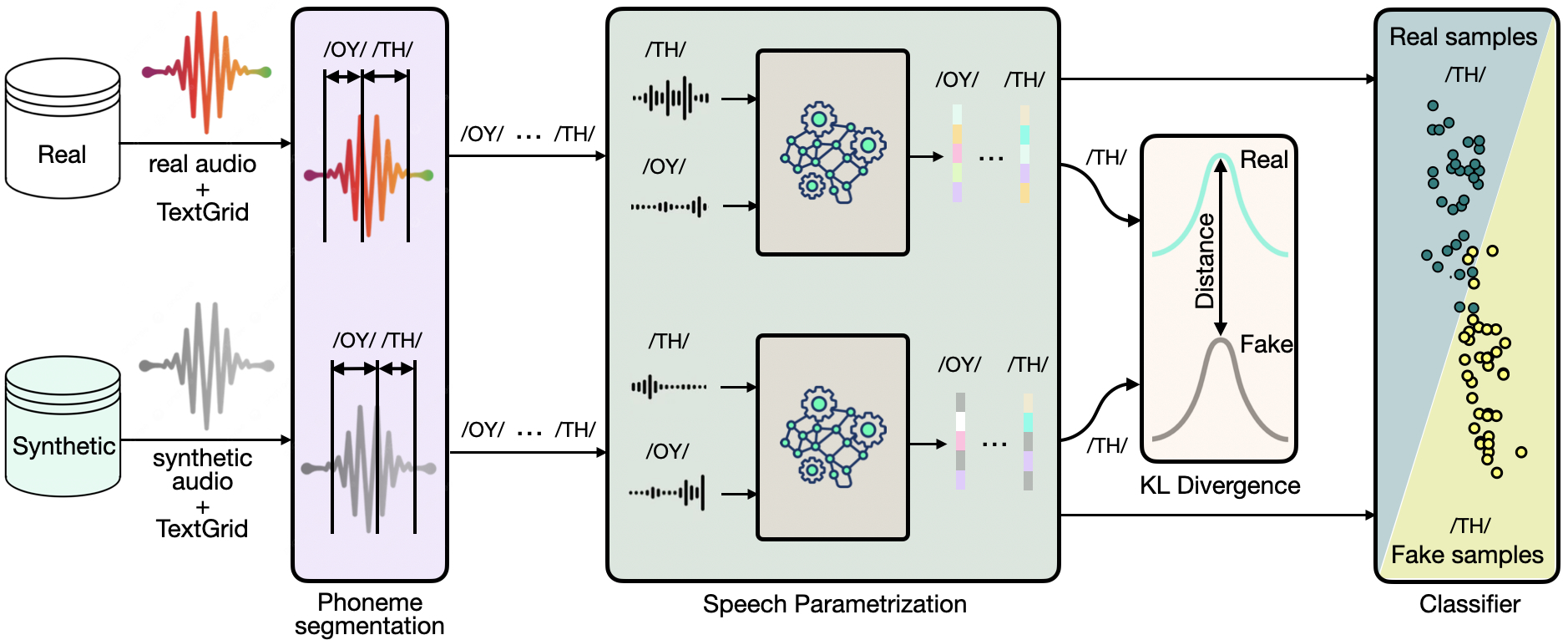}
\caption{Per-phoneme evaluation pipeline for real vs.\ synthetic discrimination.}
\label{fig:block}
\end{figure*}

All LibriSpeech files within the 100-hour subset were converted to WAV format, and their associated text transcripts were separated to ensure consistent filename alignment between text and audio. The TTS models synthesized speech directly from the transcripts, whereas the VC models used the original real audio as input and generated converted versions, as illustrated in Figure~\ref{fig:pipeline1}. When built-in reference speakers were available in a model, we utilized reference speakers directly. Ten speakers (five males and five females) from the VCTK corpus~\citep{VCTK} were selected as reference speakers to ensure consistent speaker identity and acoustic diversity across the generated data. As MeloTTS and VITS systems provide built-in voices, all available voices were used directly. For systems without built-in voices, the selected VCTK reference speaker recordings were used to condition the generated speech. We did not attempt to clone the original LibriSpeech speakers. All synthesized files were resampled to 16~kHz to maintain a uniform sampling rate across the systems.

\subsection{Phoneme-level Segmentation}

We relied on the Montreal Forced Aligner (MFA) \citep{McAuliffe2017MFA} to generate phoneme boundaries aligned with transcripts. Alignment was performed using the pretrained American English ARPAbet acoustic model (english\_us\_arpa) and its pronunciation dictionary. ARPAbet was chosen as it is the representation adopted for LibriSpeech and MFA. Thus, it ensures compatibility with alignment boundaries and the CMU pronunciation lexicon, while avoiding additional phoneme mapping or Grapheme-to-Phoneme (G2P) inference. To reduce sparsity and focus on phoneme-level acoustic characteristics, stress markers were removed (AA0/AA1/AA2 → AA). MFA then applies Viterbi forced alignment to produce TextGrid files containing phoneme labels and precise temporal boundaries. Table~\ref{tab:dataset_combined} summarizes the overall statistics of the PhonemeDF dataset. In total, it comprises approximately 730 hours of speech, equivalent to 199{,}773 synthetic speech samples, along with their respective TextGrid files, derived from 28,539 (100 hours) real utterances from the LibriSpeech corpus. Selected reference audios and their synthetic counterparts are available at the following link\footnote{https://github.com/Vamshi-Nallaguntla/PhonemeDF}. Note that a handful of TextGrids were inspected to verify alignment quality across both real and synthetic recordings.

\begin{table}
\centering
\small
\begin{tabularx}{\columnwidth}{lccc}
\hline
\textbf{Dataset} & \textbf{\# Files} & \textbf{Avg. (s)} & \textbf{Total (h)} \\
\hline
LibriSpeech  & 28,539  & 12.69 & 100 \\
MeloTTS              & 28,539  & 10.15 & 80 \\
XTTS v2              & 28,539  & 10.47 & 82 \\
Chatterbox TTS       & 28,539  & 10.15 & 80 \\
VITS TTS             & 28,539  & 10.77 & 85 \\
Chatterbox VC        & 28,539  & 12.70 & 100 \\
FreeVC               & 28,539  & 12.68 & 100 \\
StarGAN VC           & 28,539  & 12.68 & 100 \\
\hline
PhonemeDF & 199,773 & 11.37 & $\approx$ 730 \\
\hline
\end{tabularx}
\caption{\label{tab:dataset_combined} Statistics showing number of files, average duration of an audio file (in seconds), and total duration (in hours) for real and synthetic datasets.}
\end{table}

\section{Experimental Setup}
\label{setup}
In our experiments, we assess the capability of distinct synthesizers to generate synthetic phonemes. Our hypothesis is that certain phonemes are more challenging to generate, and therefore, classifying those phonemes will lead to higher accuracies compared to phonemes that are easier to generate and therefore more similar to their real counterparts. In the following section, we describe how we use PhonemeDF for the purpose of this study.

\subsection{Phoneme-Level Evaluation Setup}

\begin{figure}
    \centering
    \includegraphics[width=0.49\textwidth]{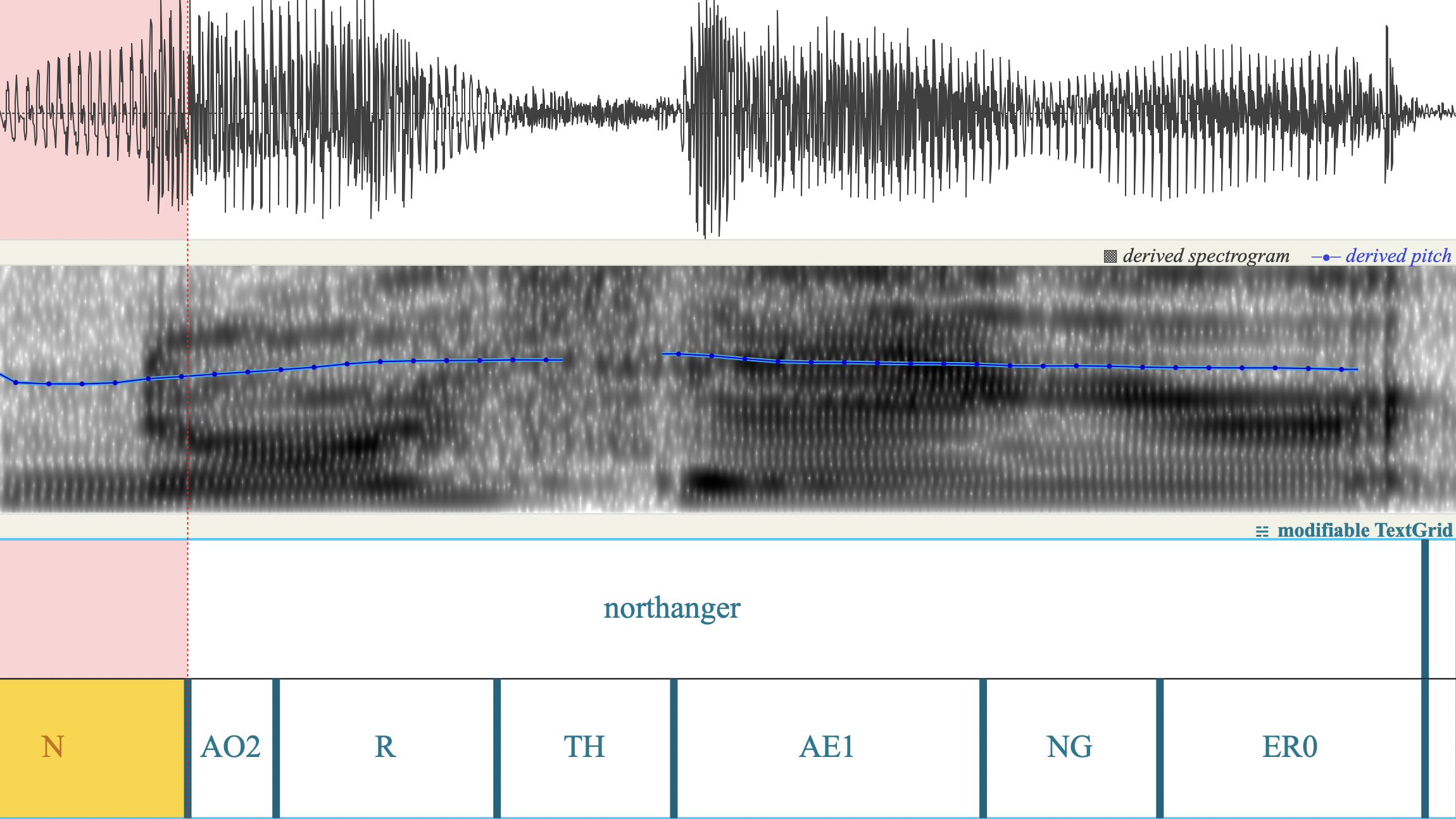}
    \caption{TextGrid showing phoneme boundaries aligned using Montreal Forced Aligner (MFA). }
    \label{fig:textgrid}
\end{figure}

As shown in Figure~\ref{fig:block}, we conduct our evaluation using parallel samples comprising real and synthetic speech. PhonemeDF includes the full 100 hours of real speech from the LibriSpeech subset together with their corresponding TextGrid alignments. Synthetic versions of these recordings are generated for each synthesis system to form parallel real–synthetic pairs. For computational efficiency, the experiments were conducted on a controlled subset of the dataset. Specifically, we selected 1000 real utterances from the LibriSpeech portion of the corpus and used their corresponding synthetic counterparts from each system. The subset was constructed in a balanced manner to ensure that all 251 speakers from the LibriSpeech subset are represented. Using phoneme boundaries extracted from the TextGrid files, speech signals are segmented into phoneme-level audio units. Figure~\ref{fig:textgrid} shows a sample TextGrid with phoneme and word boundaries. Each phoneme segment is processed separately and undergoes speech parametrization. We adopt both handcrafted and deep learning speech representations with the aim of assessing how synthetic phonemes behave within these categories. Thus, for handcrafted features, we rely on log-Mel Spectrograms (LogSpec) and Linear-Frequency Cepstral Coefficients (LFCC), and for SSL embeddings we considered WavLM~\citep{chen2022wavlm} and wav2vec 2.0~\citep{baevski2020wav2vec}.

\begin{figure*}
\centering
\begin{subfigure}[t]{.9\columnwidth}
    \includegraphics[width=\linewidth]{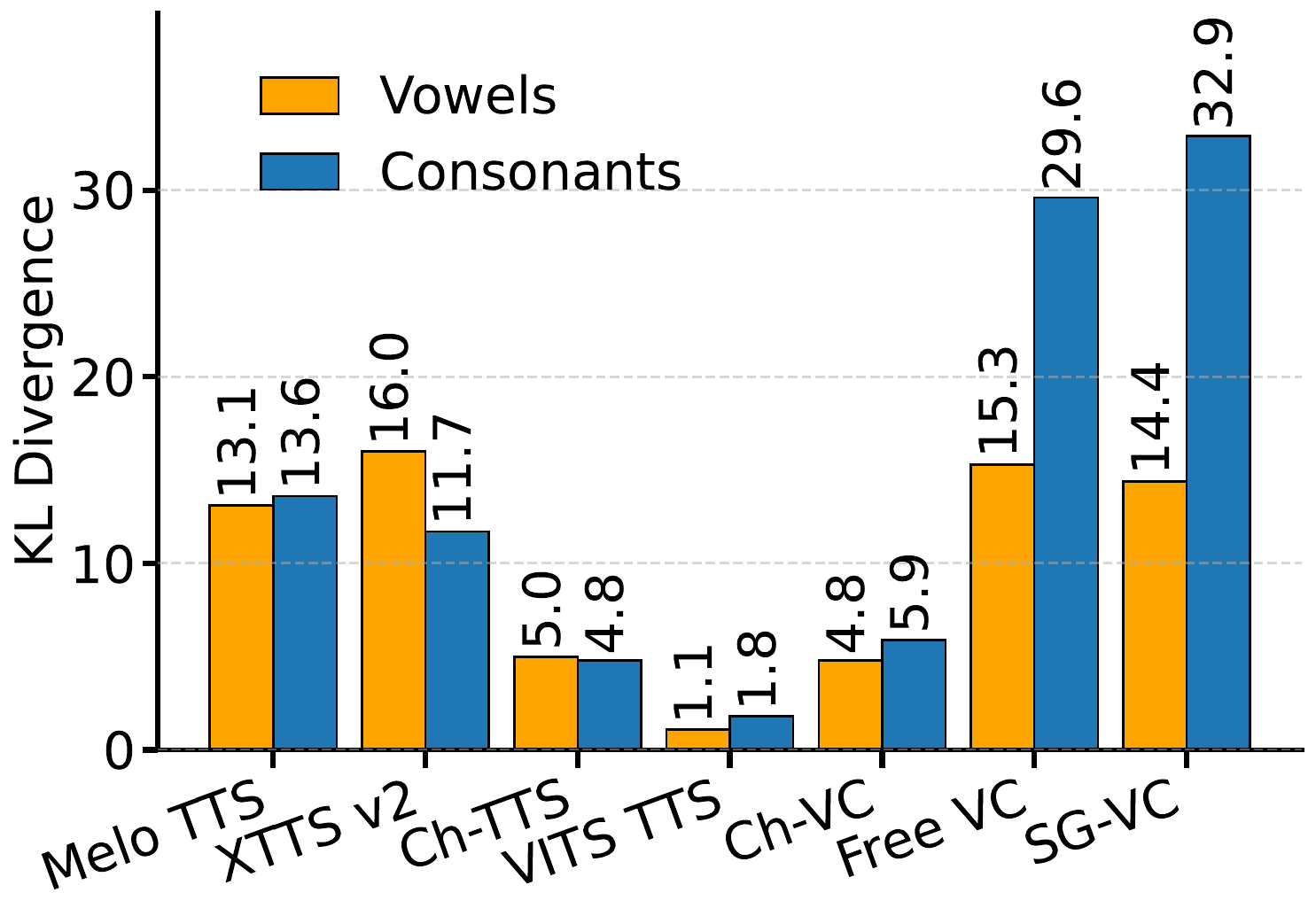}
    \caption{LFCC}
    \label{fig:lfcc_kld_sub}
\end{subfigure}
\hfill
\begin{subfigure}[t]{.9\columnwidth}
    \includegraphics[width=\linewidth]{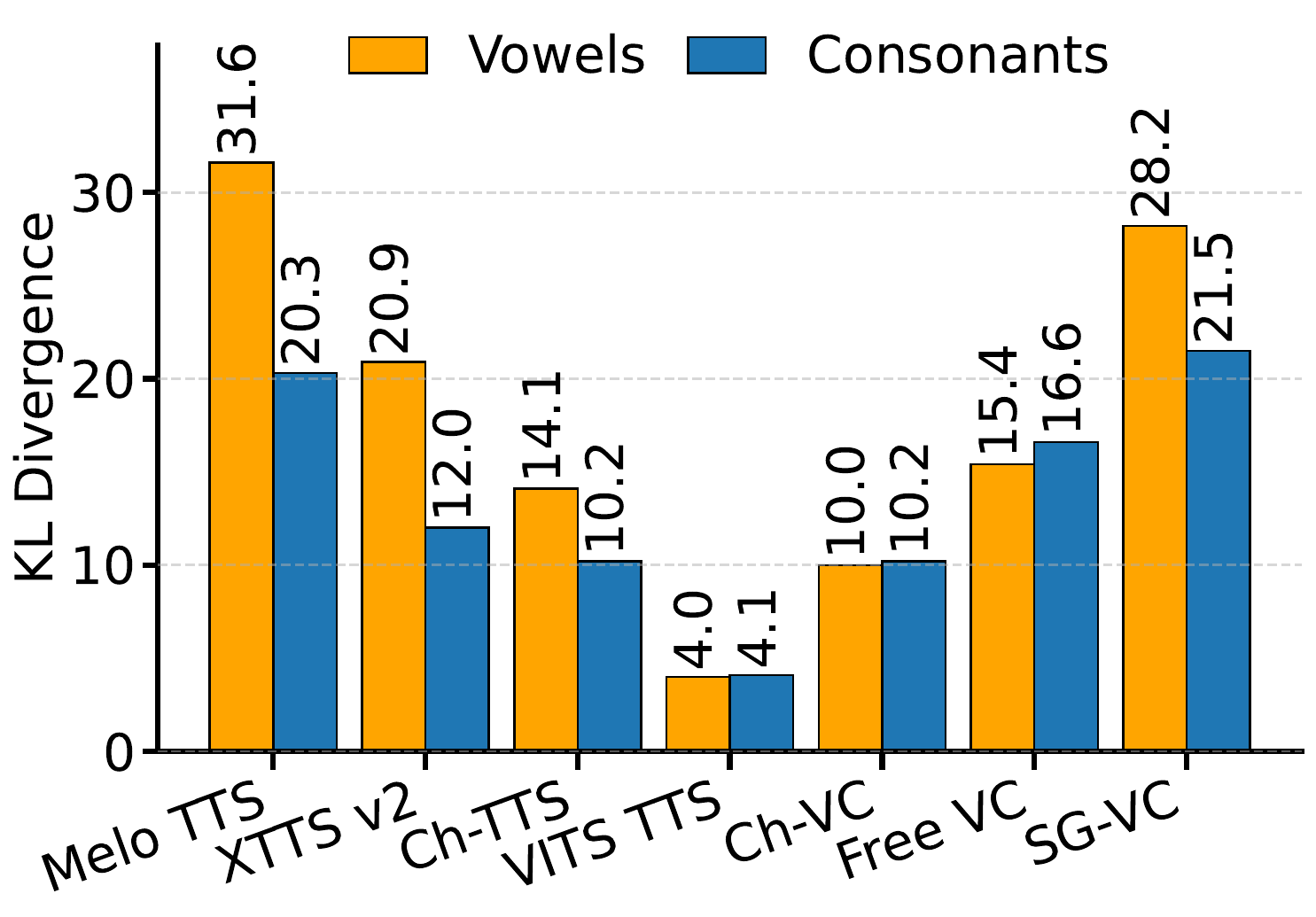}
    \caption{LogSpec}
    \label{fig:logspec_kld_sub}
\end{subfigure}

\caption{KLD between synthesized and real speech using handcrafted features.}
\label{fig:LFCC_LOGSPEC}
\end{figure*}

\begin{figure*}
\centering
\begin{subfigure}[t]{.9\columnwidth}
    \includegraphics[width=\linewidth]{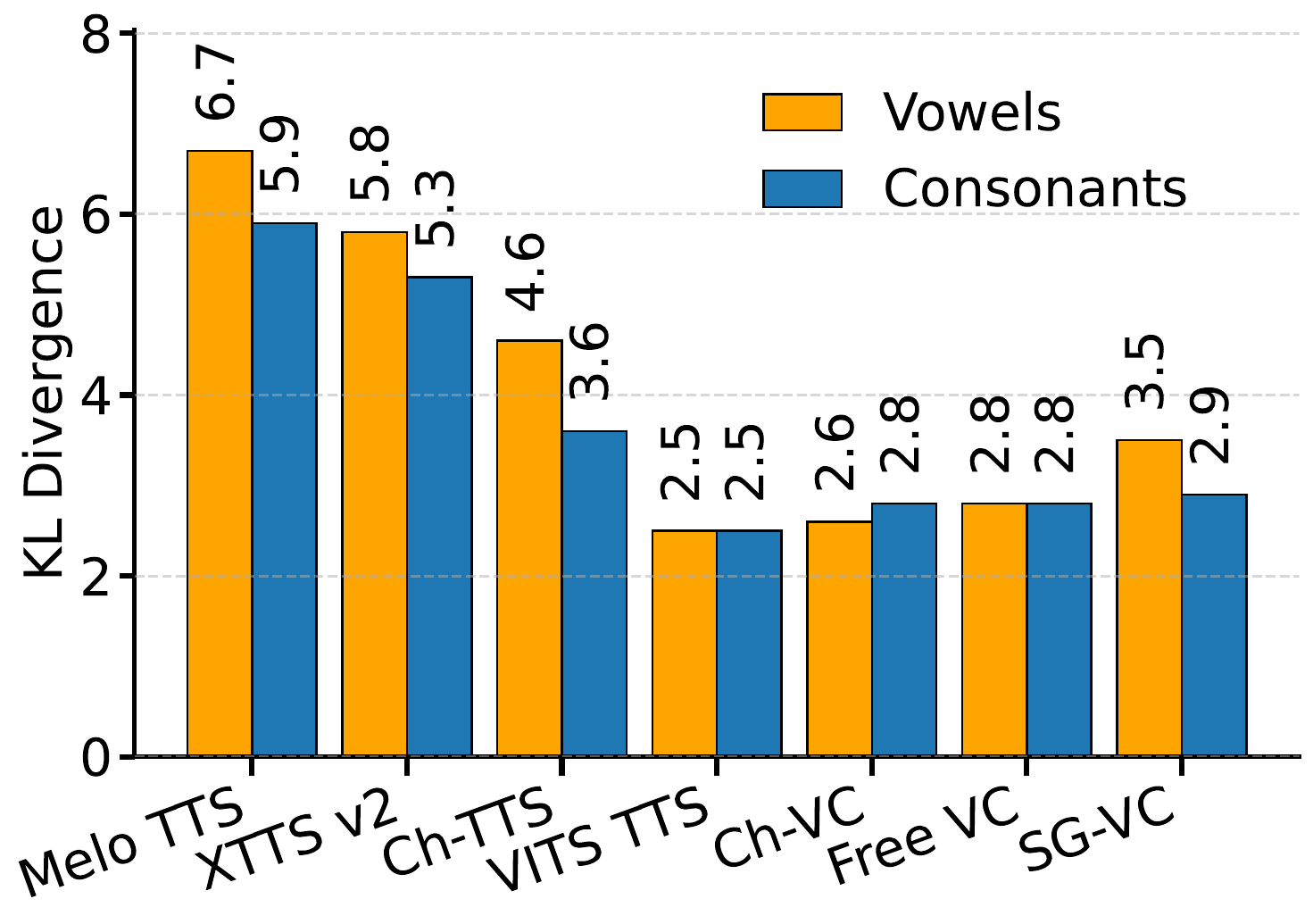}
    \caption{Wav2vec2}
    \label{fig:w2v2_kld_sub}
\end{subfigure}
\hfill
\begin{subfigure}[t]{.9\columnwidth}
    \includegraphics[width=\linewidth]{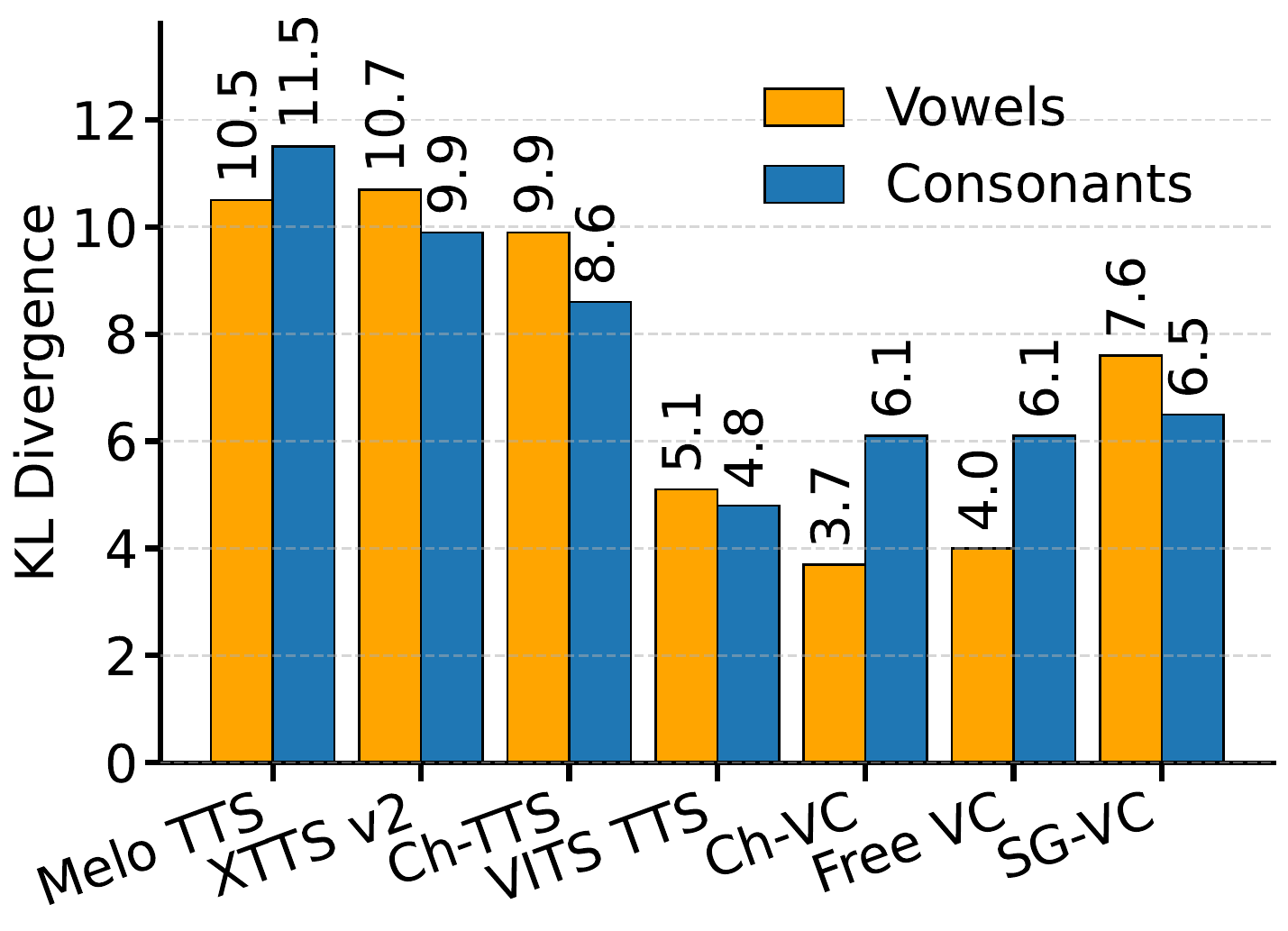}
    \caption{WavLM}
    \label{fig:wavlm_kld_sub}
\end{subfigure}

\caption{KLD between synthesized and real speech using self-supervised features.}
\label{fig:WAV2VEC_WAVLM}
\end{figure*}

For the handcrafted features, LogSpec was computed using 80 Mel bins, a 512-point Fast Fourier Transform (FFT), and a 10~ms hop size, while LFCCs used 20 coefficients with identical parameters. Frame-level features were then mean-pooled across each phoneme segment using the boundary timestamps from MFA TextGrids. Similarly, for the SSL representations, phoneme-level embeddings were extracted based on frame-level hidden states, and mean pooling was applied to each phoneme segment. Both models process audio at a 16~kHz sampling rate with a 25~ms window size, with a 20~ms hop length, producing 1024-dimensional embeddings per frame.

\subsection{Similarity of Synthetic and Real Phonemes}
To measure the similarity between real and synthetic phoneme representations, we compute the KLD between their distributions. For each phoneme in PhonemeDF, the embeddings extracted from real and synthetic speech are modeled as multivariate Gaussian distributions. Let the distributions of real and synthetic phoneme embeddings be denoted as $\mathcal{N}_R(\mu_R,\Sigma_R)$ and $\mathcal{N}_S(\mu_S,\Sigma_S)$, where $\mu$ represents the mean vector and $\Sigma$ the covariance matrix.

The KLD between two multivariate Gaussian distributions is defined as

\begin{equation}
D_{\mathrm{KL}}(\mathcal{N}_R \| \mathcal{N}_S)
\end{equation}

Since KLD is asymmetric, we compute the symmetric KLD between real and synthetic phoneme distributions:

\begin{equation}
D_{\mathrm{sym}} =
\frac{1}{2}
\left[
D_{\mathrm{KL}}(\mathcal{N}_R \| \mathcal{N}_S)
+
D_{\mathrm{KL}}(\mathcal{N}_S \| \mathcal{N}_R)
\right]
\label{eq:symkld}
\end{equation}

Higher $D_{\mathrm{sym}}$ values indicate larger distributional differences between real and synthetic phoneme embeddings, while lower values suggest higher similarity between the two distributions. We hypothesize that phonemes with higher divergence values will be easier to distinguish from their synthetic counterparts and, therefore, will yield higher classification accuracies.

\subsection{Classification Model}

Two binary classifiers are adopted to evaluate the impact of phonemes on deepfake detection. We rely on LR ~\citep{hosmer2013applied} and SVM ~\citep{cortes1995support}. This yields 78 independent classifiers per feature type and synthesis system (39~phonemes $\times$ 2~classifiers). Classification accuracy quantifies the practical detectability of each phoneme type, complementing the statistical separability measured by KLD.

As evaluation metrics, we adopted KLD, accuracy, and Pearson correlation. Pearson correlation is used to assess the relationship between phoneme-level KLD values and classification accuracy across phonemes. The null hypothesis assumes no linear relationship between the two variables, $H_0: r = 0$, where $r$ denotes the Pearson correlation coefficient.

\begin{figure*}
\centering
\begin{subfigure}[t]{.9\columnwidth}
    \includegraphics[width=\linewidth]{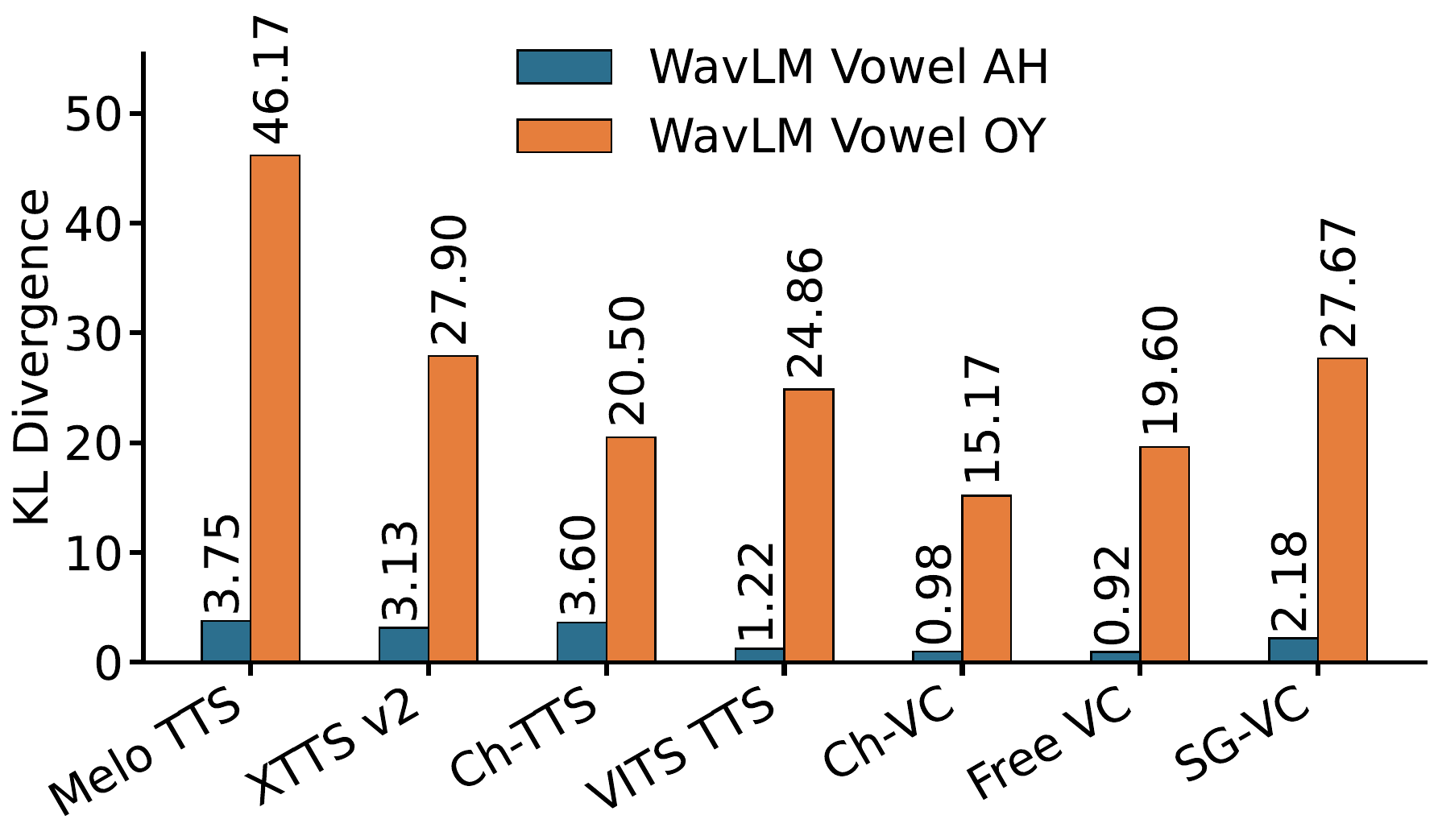}
    \caption{WavLM (Vowel)}
    \label{fig:wavlm_vowel_sub}
\end{subfigure}
\hfill
\begin{subfigure}[t]{.9\columnwidth}
    \includegraphics[width=\linewidth]{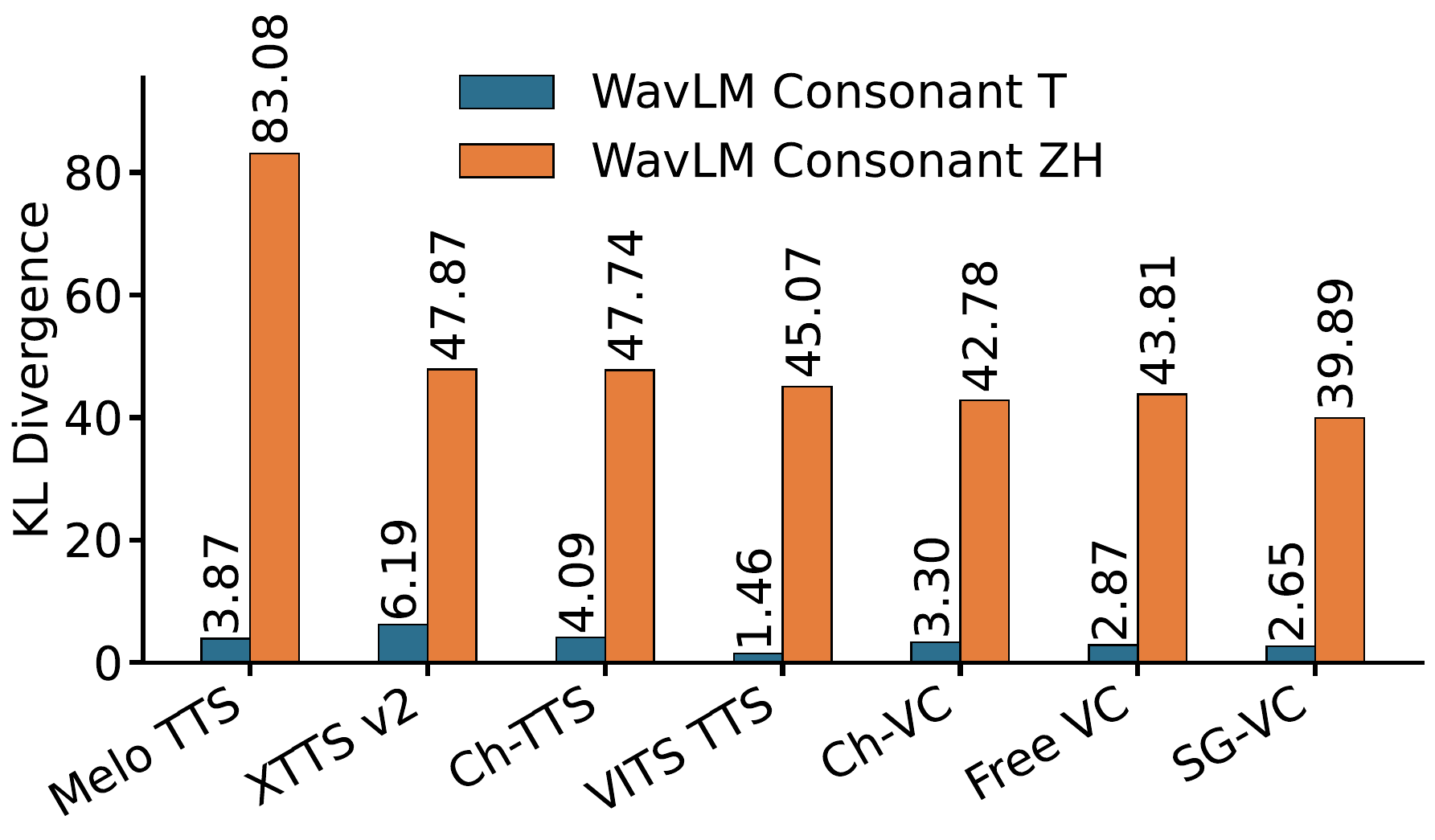}
    \caption{WavLM (Consonant)}
    \label{fig:wavlm_consonant_sub}
\end{subfigure}

\caption{Lowest and highest KLD scores for (a) WavLM vowel representation and (b) WavLM consonant representation.}
\label{fig:kldphoneme}
\end{figure*}

\section{Results and Discussion}
\label{results}

\subsection{KLD for Vowels and Consonants}

Figure~\ref{fig:LFCC_LOGSPEC} and Figure~\ref{fig:WAV2VEC_WAVLM} show the KLD scores averaged across vowels and consonants for handcrafted features and SSL embeddings. The results are presented for each synthesizer. For handcrafted representations, the divergence patterns are mostly influenced by the spectral envelope and energy differences. LFCC and LogSpec representations rank voice-conversion systems as providing the highest mismatch between synthetic and real phonemes. In particular, SG-VC produces large divergences for both vowels and consonants in LFCC, reaching KLD values of 14.4 and 32.9, respectively, while FreeVC also shows a large consonant divergence of 29.6. Similarly, LogSpec reveals substantial mismatches for MeloTTS vowels, 31.6, and SG-VC vowels, 28.2, with consonant divergence remaining high for SG-VC, 21.5, and MeloTTS, 20.3. In contrast, VITS TTS produces synthetic phonemes that are very close to real ones in the handcrafted feature space, with very low KLD values, i.e., 1.1 and 1.8 in LFCC, and 4.0 and 4.1 for LogSpec, respectively, for vowels and consonants. Overall, consonants tend to diverge more than vowels for the VC systems (e.g., FreeVC and SG-VC), reflecting the sensitivity of spectral features to transient and fricative energy patterns. For TTS systems, the behavior is more mixed. XTTS v2 and MeloTTS show larger vowel mismatches, whereas VITS exhibits slightly higher divergence for consonants. In general, LogSpec produces higher KLD scores than LFCC, suggesting that the handcrafted feature representation adopted has a significant impact on the way real and synthetic phonemes are represented. The TTS and VC systems play an equally important role.

For the SSL representations, the average KLD scores decrease considerably compared to handcrafted features. Wav2Vec2 produces moderate divergences, with MeloTTS reaching the highest values, 6.7 for vowels and 5.9 for consonants, while VITS remains closest to real speech, i.e., 2.5 for both vowels and consonants. Compared to handcrafted features, WavLM representations produce lower KLD values overall, with most systems falling between 3 and 11. In this representation, Ch-VC and FreeVC produce relatively small vowel divergences, 3.7 and 4.0, but slightly larger consonant divergences, around 6.1. These results suggest that SSL embeddings capture higher-level phonetic structures and are less sensitive to raw spectral artifacts compared to handcrafted features.

Figure~\ref{fig:kldphoneme} presents examples of individual phonemes illustrating the lowest and highest divergences observed in the WavLM representation. For vowels, the monophthong AH consistently shows very small divergence across all systems, with values between 0.92 and 3.75, whereas the diphthong OY exhibits substantially larger divergence, reaching 46.17 for MeloTTS and remaining above 15 for all systems. This indicates that dynamic vowels with complex articulatory transitions are more difficult for synthesis models to reproduce accurately. A similar pattern is observed for consonants. The stop consonant /T/ shows relatively small divergence across systems, ranging from 1.46 to 6.19, indicating that simple closure–release patterns are reproduced reliably. In contrast, the fricative /ZH/ exhibits very large divergence, reaching 83.08 for MeloTTS and remaining above 39 for all systems, suggesting that high-frequency fricative noise is difficult for both TTS and VC models to replicate. Overall, these results highlight phoneme-level variability in synthetic speech. While phonemes such as /AH/ and /T/ are reproduced with high fidelity, others such as /OY/ and /ZH/ show substantial divergence, revealing limitations of current synthesizers.

\subsection{Correlation Between KLD and ACC}

In this experiment, we present the correlations between KLD scores and the accuracies for audio deepfake detection under different configurations. Within each table, results are presented by synthesizer system, type of classifier, speech representation, and the category of phonemes: vowels or consonants. In Table~\ref{tab:kac_lfcc_vow}, we present performance for vowels with LFCC based on the LR classifier, providing moderate results for all systems, including Ch-VC, SG-VC, and Ch-TTS, where correlations range from 0.68, 0.74, and 0.79, respectively. Higher correlations are attained for consonants, as shown in Table~\ref{tab:kac_lfcc_cons}, suggesting that for the LFCC representation, classifiers should rely more on consonants to distinguish between real and synthetic audio. LogSpec features show a similar pattern, with lower correlations for specific synthesizers. Table~\ref{tab:kac_ls_vowel} provides the performance for vowels, with the highest correlations ranging from 0.84, 0.87, and 0.89, respectively, for Free VC, Melo TTS, and SG-VC. These results are based on LR, but a similar trend is found for the SVM. Low performance, however, was observed with the VITS synthesizer, with correlations as low as 0.07 and 0.04 for LR and SVM for the experiment with vowels, and a similar trend was found for consonants. These results suggest that for handcrafted features, larger KLD leads to higher classification accuracies and, therefore, phonemes that deviate more from real speech yield better detection rates. 

For SSL representations, correlations are generally lower and more inconsistent across systems. Tables~\ref{tab:kac_w2v2_vowel} and~\ref{tab:kac_w2v2_cons} present the results using Wav2Vec2 representations. In contrast to the handcrafted features, correlations vary substantially across synthesizers and phoneme categories. For vowels, in Table~\ref{tab:kac_w2v2_vowel}, several systems exhibit weak or even negative correlations, such as XTTS v2 and Ch-TTS, while VITS TTS shows a relatively high positive correlation (i.e., 0.75 for both LR and SVM). Other systems, including Ch-VC and Free VC, display moderate negative correlations, indicating that larger KLD does not necessarily translate into better classification performance when using this representation. For consonants, in Table~\ref{tab:kac_w2v2_cons}, correlations tend to be negative for most systems, particularly for Melo TTS, VITS TTS, and SG-VC, suggesting that higher divergence is often associated with lower detection accuracy. A similar pattern is observed with WavLM representations, as shown in Tables~\ref{tab:kac_wavlm_vowel} and~\ref{tab:kac_wavlm_cons}. For vowels (see Table~\ref{tab:kac_wavlm_vowel}), correlations are generally weak and fluctuate around zero for most synthesizers, with only Ch-VC and Free VC showing moderate negative correlations. For consonants (Table~\ref{tab:kac_wavlm_cons}), however, consistently strong negative correlations are observed across nearly all systems, with coefficients reaching as low as -0.83 for SG-VC and -0.79 for Melo TTS using LR. This indicates that, unlike handcrafted features, larger KLD in SSL representations often corresponds to lower classification accuracy. Overall, these results suggest that the relationship between phoneme-level divergence and detection performance differs substantially between handcrafted and SSL representations, with the latter exhibiting weaker and frequently inverse correlations.

\begin{table}[!ht]
\centering
\small
\setlength{\tabcolsep}{5pt}
\renewcommand{\arraystretch}{1.1}
\begin{tabular}{lcccc}
\toprule
& \multicolumn{2}{c}{\textbf{LR}} & \multicolumn{2}{c}{\textbf{SVM}}\\
\cmidrule(lr){2-3} \cmidrule(lr){4-5} 
\textbf{System} & $r$ & $p$-value & $r$ & $p$-value\\
\midrule

Melo TTS & 0.68 & 4.99E-03 & 0.63 & 1.16E-02 \\
XTTS v2 & 0.67 & 6.11E-03 & 0.69 & 4.07E-03 \\
Ch-TTS & 0.79 & 4.94E-04 & 0.80 & 3.89E-04 \\
VITS TTS & 0.49 & 0.065569 & 0.51 & 0.054600 \\
Ch-VC & 0.68 & 5.38E-03 & 0.69 & 4.05E-03 \\
Free VC & 0.71 & 3.06E-03 & 0.71 & 2.89E-03 \\
SG-VC & 0.74 & 1.73E-03 & 0.73 & 1.97E-03 \\

\bottomrule
\end{tabular}
\caption{Results for vowels based on LFCC representation.}
\label{tab:kac_lfcc_vow}
\end{table}

\begin{table}[!ht]
\centering
\small
\setlength{\tabcolsep}{5pt}
\renewcommand{\arraystretch}{1.1}
\begin{tabular}{lcccc}
\toprule
& \multicolumn{2}{c}{\textbf{LR}} & \multicolumn{2}{c}{\textbf{SVM}}\\
\cmidrule(lr){2-3} \cmidrule(lr){4-5} 
\textbf{System} & $r$ & $p$-value & $r$ & $p$-value\\
\midrule

Melo TTS & 0.79 & 4.28E-06 & 0.78 & 5.99E-06 \\
XTTS v2 & 0.76 & 1.47E-05 & 0.76 & 1.69E-05 \\
Ch-TTS & 0.85 & 1.11E-07 & 0.85 & 1.38E-07 \\
VITS TTS & 0.65 & 5.79E-04 & 0.66 & 5.08E-04 \\
Ch-VC & 0.88 & 1.22E-08 & 0.84 & 2.92E-07 \\
Free VC & 0.65 & 6.37E-04 & 0.66 & 5.12E-04 \\
SG-VC & 0.82 & 1.16E-06 & 0.82 & 1.21E-06 \\

\bottomrule
\end{tabular}
\caption{Results for consonants based on LFCC representation.}
\label{tab:kac_lfcc_cons}
\end{table}

\begin{table}[!ht]
\centering
\small
\setlength{\tabcolsep}{5pt}
\renewcommand{\arraystretch}{1.1}
\begin{tabular}{lcccc}
\toprule
& \multicolumn{2}{c}{\textbf{LR}} & \multicolumn{2}{c}{\textbf{SVM}}\\
\cmidrule(lr){2-3} \cmidrule(lr){4-5} 
\textbf{System} & $r$ & $p$-value & $r$ & $p$-value\\
\midrule

Melo TTS & 0.29 & 0.298568 & 0.27 & 0.338104 \\
XTTS v2 & -0.43 & 0.109768 & -0.39 & 0.155521 \\
Ch-TTS & -0.36 & 0.187258 & -0.50 & 0.059611 \\
VITS TTS & 0.75 & 1.30E-03 & 0.75 & 1.39E-03 \\
Ch-VC & -0.60 & 0.017691 & -0.61 & 0.016767 \\
Free VC & -0.73 & 2.13E-03 & -0.79 & 5.05E-04 \\
SG-VC & 0.33 & 0.227209 & 0.21 & 0.456768 \\

\bottomrule
\end{tabular}
\caption{Results for vowels based on Wav2vec 2.0 representation.}
\label{tab:kac_w2v2_vowel}
\end{table}

\begin{table}[!ht]
\centering
\small
\setlength{\tabcolsep}{5pt}
\renewcommand{\arraystretch}{1.1}
\begin{tabular}{lcccc}
\toprule
& \multicolumn{2}{c}{\textbf{LR}} & \multicolumn{2}{c}{\textbf{SVM}}\\
\cmidrule(lr){2-3} \cmidrule(lr){4-5} 
\textbf{System} & $r$ & $p$-value & $r$ & $p$-value\\
\midrule

Melo TTS & -0.50 & 0.013894 & -0.59 & 2.19E-03 \\
XTTS v2 & 0.31 & 0.134920 & 0.02 & 0.908695 \\
Ch-TTS & -0.43 & 0.036489 & -0.25 & 0.232014 \\
VITS TTS & -0.56 & 4.25E-03 & -0.16 & 0.443928 \\
Ch-VC & -0.38 & 0.069623 & -0.35 & 0.090622 \\
Free VC & -0.29 & 0.173695 & -0.01 & 0.963936 \\
SG-VC & -0.59 & 2.61E-03 & -0.43 & 0.038125 \\

\bottomrule
\end{tabular}
\caption{Results for consonants based on Wav2vec 2.0 representation.}
\label{tab:kac_w2v2_cons}
\end{table}

\begin{table}[!ht]
\centering
\small
\setlength{\tabcolsep}{5pt}
\renewcommand{\arraystretch}{1.1}
\begin{tabular}{lcccc}
\toprule
& \multicolumn{2}{c}{\textbf{LR}} & \multicolumn{2}{c}{\textbf{SVM}}\\
\cmidrule(lr){2-3} \cmidrule(lr){4-5} 
\textbf{System} & $r$ & $p$-value & $r$ & $p$-value\\
\midrule

Melo TTS & 0.87 & 2.48E-05 & 0.81 & 2.41E-04 \\
XTTS v2 & 0.70 & 3.55E-03 & 0.76 & 9.30E-04 \\
Ch-TTS & 0.65 & 9.37E-03 & 0.74 & 1.59E-03 \\
VITS TTS & 0.07 & 0.803638 & 0.04 & 0.898342 \\
Ch-VC & 0.52 & 0.046131 & 0.61 & 0.014760 \\
Free VC & 0.84 & 7.92E-05 & 0.88 & 1.69E-05 \\
SG-VC & 0.89 & 8.82E-06 & 0.91 & 3.19E-06 \\

\bottomrule
\end{tabular}
\caption{Results for vowels based on LogSpec representation.}
\label{tab:kac_ls_vowel}
\end{table}

\begin{table}[!ht]
\centering
\small
\setlength{\tabcolsep}{5pt}
\renewcommand{\arraystretch}{1.1}
\begin{tabular}{lcccc}
\toprule
& \multicolumn{2}{c}{\textbf{LR}} & \multicolumn{2}{c}{\textbf{SVM}}\\
\cmidrule(lr){2-3} \cmidrule(lr){4-5} 
\textbf{System} & $r$ & $p$-value & $r$ & $p$-value\\
\midrule

Melo TTS & 0.41 & 0.046255 & 0.56 & 4.82E-03 \\
XTTS v2 & 0.62 & 1.29E-03 & 0.55 & 4.94E-03 \\
Ch-TTS & 0.57 & 3.70E-03 & 0.54 & 6.44E-03 \\
VITS TTS & 0.00 & 0.997485 & -0.17 & 0.428621 \\
Ch-VC & 0.74 & 3.57E-05 & 0.69 & 1.86E-04 \\
Free VC & 0.61 & 1.72E-03 & 0.63 & 9.98E-04 \\
SG-VC & 0.63 & 8.65E-04 & 0.71 & 1.04E-04 \\

\bottomrule
\end{tabular}
\caption{Results for consonants based on LogSpec representation.}
\label{tab:kac_ls_cons}
\end{table}

\begin{table}[!ht]
\centering
\small
\setlength{\tabcolsep}{5pt}
\renewcommand{\arraystretch}{1.1}
\begin{tabular}{lcccc}
\toprule
 & \multicolumn{2}{c}{\textbf{LR}} & \multicolumn{2}{c}{\textbf{SVM}}\\
\cmidrule(lr){2-3} \cmidrule(lr){4-5} 
\textbf{System} & $r$ & $p$-value & $r$ & $p$-value\\
\midrule

Melo TTS & -0.41 & 0.132823 & -0.31 & 0.264627 \\
XTTS v2 & 0.18 & 0.527040 & 0.17 & 0.537221 \\
Ch-TTS & -0.04 & 0.882777 & 0.01 & 0.965355 \\
VITS TTS & 0.40 & 0.145025 & 0.47 & 0.080362 \\
Ch-VC & -0.65 & 8.13E-03 & -0.67 & 6.15E-03 \\
Free VC & -0.78 & 6.69E-04 & -0.78 & 6.43E-04 \\
SG-VC & -0.10 & 0.718714 & -0.21 & 0.453530 \\

\bottomrule
\end{tabular}
\caption{Results for vowels based on WavLM representation.}
\label{tab:kac_wavlm_vowel}
\end{table}

\begin{table}[!ht]
\centering
\small
\setlength{\tabcolsep}{5pt}
\renewcommand{\arraystretch}{1.1}
\begin{tabular}{lcccc}
\toprule
& \multicolumn{2}{c}{\textbf{LR}} & \multicolumn{2}{c}{\textbf{SVM}}\\
\cmidrule(lr){2-3} \cmidrule(lr){4-5} 
\textbf{System} & $r$ & $p$-value & $r$ & $p$-value\\
\midrule

Melo TTS & -0.79 & 4.18E-06 & -0.74 & 3.43E-05 \\
XTTS v2 & -0.60 & 1.93E-03 & -0.60 & 1.89E-03 \\
Ch-TTS & -0.56 & 4.76E-03 & -0.61 & 1.44E-03 \\
VITS TTS & -0.78 & 7.42E-06 & -0.73 & 5.40E-05 \\
Ch-VC & -0.58 & 3.19E-03 & -0.53 & 8.13E-03 \\
Free VC & -0.33 & 0.115046 & -0.44 & 0.032071 \\
SG-VC & -0.80 & 2.82E-06 & -0.75 & 2.55E-05 \\

\bottomrule
\end{tabular}
\caption{Results for consonants based on WavLM representation.}
\label{tab:kac_wavlm_cons}
\end{table}

\subsection{Phoneme Rankings Across Systems}

We identify the most discriminative phoneme categories by analyzing per-phoneme KLD, LR accuracy, and SVM accuracy across all feature representations. Across systems and features, consonants generally exhibit stronger discriminability than vowels. Handcrafted spectral representations highlight this difference clearly: LogSpec produces the highest divergence values (average KLD of 31.6 for vowels and 20.3 for consonants), followed by LFCC (13.1 for vowels and 13.6 for consonants), while SSL embeddings yield lower divergence (e.g., WavLM averages of 10.5 for vowels and 11.5 for consonants). These results indicate that handcrafted spectral features emphasize acoustic mismatches between real and synthetic speech more strongly, whereas SSL embeddings capture more subtle phonetic variations. Among vowels, diphthongs consistently appear as the most discriminative category across feature types. In particular, /OY/ and /EY/ frequently exhibit the highest divergence values across synthesis systems, followed by /AW/ and /AY/. These phonemes involve dynamic formant trajectories that are difficult for generative models to reproduce accurately. In contrast, simpler monophthongs such as /AH/ and /UH/ consistently show lower divergence values, indicating that they are easier for synthesis systems to approximate. For consonants, fricatives and plosives dominate the top ranks. Fricatives such as /SH/, /S/, and /ZH/ exhibit large divergence values in spectral features due to their broadband noise characteristics, while plosives including /P/, /B/, and /T/ frequently appear among the most discriminative phonemes in SSL embeddings, reflecting the importance of transient temporal cues captured by these models. When considering vowels and consonants jointly, three phoneme groups consistently emerge as discriminative across systems and feature types: diphthongs (e.g., /OY/), fricatives (e.g., /SH/ and /S/), and plosives (e.g., /P/ and /B/). These categories capture different synthesis artifacts, including complex formant transitions, sustained spectral turbulence, and rapid articulatory bursts.

Our findings align with and extend recent phoneme-level deepfake analyses. Temmar et al.~\citep{temmar2025phonetic} also showed that diphthongs and fricatives are highly discriminative, consistent with our observation that /OY/, /SH/, and /F/ frequently appear among the most informative phonemes. However, we extend prior work by evaluating 39 phoneme categories across seven synthesis systems and four feature representations. Different feature types reveal complementary phonetic sensitivities. SSL embeddings (WavLM and wav2vec~2.0) emphasize transient consonants such as plosives, while handcrafted spectral features (LogSpec and LFCC) highlight fricatives and diphthongs that manifest as spectral irregularities. At the system level, StarGAN-VC and MeloTTS consistently exhibit the largest divergence from real speech, while VITS remains closest to natural speech. Overall, handcrafted features highlight broader spectral mismatches, whereas SSL embeddings reveal finer phonetic inconsistencies, providing complementary insights into synthesis realism and phoneme-level variability.

\section{Conclusion}
\label{conclusion}

This work introduced PhonemeDF, a dataset for audio deepfake detection with phoneme-level annotations. It comprises nearly 200k synthetic utterances and about two million aligned phoneme segments generated using seven TTS and VC systems. We used the data to analyze the discriminability of phonemes through KLD and supervised classification using both handcrafted spectral features and SSL speech representations. Our results show that certain phoneme categories—particularly diphthongs, fricatives, and plosives—consistently provide strong cues for distinguishing synthetic from real speech. Handcrafted spectral representations emphasize large acoustic mismatches between real and synthetic speech, while SSL embeddings capture more subtle phonetic inconsistencies. Additionally, we observe systematic differences across synthesis models, with voice conversion systems generally producing larger phoneme-level deviations from natural speech than modern TTS models. Overall, our findings highlight the value of phoneme-level analysis for understanding synthesis artifacts and suggest that combining complementary feature representations may improve the robustness of future deepfake detection systems.

\section{Ethics Statement and Limitations}
While this work aims to improve the detection of synthetic speech, the dataset and analysis may also indirectly facilitate the study of synthesis artifacts that could be exploited to improve generation systems. The dataset is restricted to English speech and a limited set of synthesis models, which may limit the generalization of the findings to other languages or emerging speech generation technologies. Furthermore, the reference speech is derived from a specific corpus and recording condition, which may introduce biases in speaker characteristics, recording environments, and speaking styles. In addition, phoneme boundaries are obtained using forced alignment, which may introduce small segmentation errors that affect phoneme-level analysis. Our experiments rely on a limited set of feature representations and relatively simple classifiers, and the evaluation focuses primarily on statistical divergence and classification accuracy without extensive perceptual validation. Future work will expand the dataset to additional languages and synthesis systems and incorporate perceptual studies to better relate phoneme-level differences to human judgments. We will also investigate phoneme transitions, as artifacts at phoneme boundaries may help detect partially manipulated synthetic speech.

\section{Bibliographical References}\label{sec:reference}

\bibliographystyle{lrec2026-natbib}
\bibliography{lrec2026-example}

@article{nusbaum1997measuring,
  title={Measuring the naturalness of synthetic speech},
  author={Nusbaum, Howard C. and Francis, Alexander L. and Henly, Anne S.},
  journal={International Journal of Speech Technology},
  volume={2},
  pages={7--19},
  year={1997},
  publisher={Springer}
}

@article{ren2019fastspeech,
  title={FastSpeech: Fast, Robust and Controllable Text to Speech},
  author={Ren, Yi and Ruan, Yangjun and Tan, Xu and Qin, Tao and Zhao, Sheng and Zhao, Zhou and Liu, Tie-Yan},
  journal={Advances in Neural Information Processing Systems},
  volume={32},
  year={2019}
}

@inproceedings{prenger2019waveglow,
  title={WaveGlow: A Flow-based Generative Network for Speech Synthesis},
  author={Prenger, Ryan and Valle, Rafael and Catanzaro, Bryan},
  booktitle={ICASSP 2019 - 2019 IEEE International Conference on Acoustics, Speech and Signal Processing (ICASSP)},
  pages={3617--3621},
  year={2019},
  organization={IEEE}
}

@article{naayini2025ai,
  title={AI-Powered Assistive Technologies for Visual Impairment},
  author={Naayini, Prudhvi and Myakala, Praveen Kumar and Bura, Chiranjeevi and Jonnalagadda, Anil Kumar and Kamatala, Srikanth},
  journal={arXiv preprint arXiv:2503.15494},
  year={2025}
}

@article{yamagishi2021asvspoof,
  title={ASVspoof 2021: Accelerating Progress in Spoofed and Deepfake Speech Detection},
  author={Yamagishi, Junichi and Wang, Xin and Todisco, Massimiliano and Sahidullah, Md and Patino, Jose and Nautsch, Andreas and Liu, Xuechen and Lee, Kong Aik and Kinnunen, Tomi and Evans, Nicholas and others},
  journal={arXiv preprint arXiv:2109.00537},
  year={2021}
}

@inproceedings{wu2015asvspoof,
  title={ASVspoof 2015: The First Automatic Speaker Verification Spoofing and Countermeasures Challenge},
  author={Wu, Zhizheng and Kinnunen, Tomi and Evans, Nicholas and Yamagishi, Junichi and Hanil{\c{c}}i, Cemal and Sahidullah, Md and Sizov, Aleksandr},
  booktitle={INTERSPEECH 2015, Automatic Speaker Verification Spoofing and Countermeasures Challenge, colocated with INTERSPEECH 2015},
  pages={2037--2041},
  year={2015},
  organization={ISCA}
}

@inproceedings{yi2022add,
  title={ADD 2022: The First Audio Deep Synthesis Detection Challenge},
  author={Yi, Jiangyan and Fu, Ruibo and Tao, Jianhua and Nie, Shuai and Ma, Haoxin and Wang, Chenglong and Wang, Tao and Tian, Zhengkun and Bai, Ye and Fan, Cunhang and others},
  booktitle={ICASSP 2022 - 2022 IEEE International Conference on Acoustics, Speech and Signal Processing (ICASSP)},
  pages={9216--9220},
  year={2022},
  organization={IEEE}
}

@misc{melotts2024,
  author={Zhao, Wenliang and Yu, Xumin and Qin, Zengyi},
  title={MeloTTS: High-quality Multi-lingual Multi-accent Text-to-Speech},
  howpublished={GitHub repository},
  note={\url{https://github.com/myshell-ai/MeloTTS}},
  year={2023}
}

@article{xttsv2_2024,
  title={XTTS: A Massively Multilingual Zero-shot Text-to-Speech Model},
  author={Casanova, Edresson and Davis, Kelly and G{\"o}lge, Eren and G{\"o}knar, G{\"o}rkem and Gulea, Iulian and Hart, Logan and Aljafari, Aya and Meyer, Joshua and Morais, Reuben and Olayemi, Samuel and others},
  journal={arXiv preprint arXiv:2406.04904},
  year={2024}
}

@misc{Chatterbox2024,
  author={{Resemble AI}},
  title={{Chatterbox-TTS}},
  howpublished={GitHub repository},
  note={\url{https://github.com/resemble-ai/chatterbox}},
  year={2025}
}

@inproceedings{VITS2021,
  title={Conditional Variational Autoencoder with Adversarial Learning for End-to-End Text-to-Speech},
  author={Kim, Jaehyeon and Kong, Jungil and Son, Juhee},
  booktitle={International Conference on Machine Learning},
  pages={5530--5540},
  year={2021},
  organization={PMLR}
}

@inproceedings{FreeVC-23,
  title={FreeVC: Towards High-quality Text-free One-shot Voice Conversion},
  author={Li, Jingyi and Tu, Weiping and Xiao, Li},
  booktitle={ICASSP 2023 - 2023 IEEE International Conference on Acoustics, Speech and Signal Processing (ICASSP)},
  pages={1--5},
  year={2023},
  organization={IEEE}
}

@article{StarGANv2VC-21,
  title={StarGANv2-VC: A Diverse, Unsupervised, Non-parallel Framework for Natural-sounding Voice Conversion},
  author={Li, Yinghao Aaron and Zare, Ali and Mesgarani, Nima},
  journal={arXiv preprint arXiv:2107.10394},
  year={2021}
}

@inproceedings{suthokumar2019phoneme,
  title={Phoneme Specific Modelling and Scoring Techniques for Anti Spoofing System},
  author={Suthokumar, Gajan and Sriskandaraja, Kaavya and Sethu, Vidhyasaharan and Wijenayake, Chamith and Ambikairajah, Eliathamby},
  booktitle={ICASSP 2019 - 2019 IEEE International Conference on Acoustics, Speech and Signal Processing (ICASSP)},
  pages={6106--6110},
  year={2019},
  organization={IEEE}
}

@inproceedings{dhamyal2021using,
  title={Using Self Attention DNNs to Discover Phonemic Features for Audio Deep Fake Detection},
  author={Dhamyal, Hira and Ali, Ayesha and Qazi, Ihsan Ayyub and Raza, Agha Ali},
  booktitle={2021 IEEE Automatic Speech Recognition and Understanding Workshop (ASRU)},
  pages={1178--1184},
  year={2021},
  organization={IEEE}
}

@inproceedings{sivaraman2025voiced,
  title={Investigating Voiced and Unvoiced Regions of Speech for Audio Deepfake Detection},
  author={Sivaraman, Ganesh and Tak, Hemlata and Khoury, Elie},
  booktitle={ICASSP 2025 - 2025 IEEE International Conference on Acoustics, Speech and Signal Processing (ICASSP)},
  pages={1--5},
  year={2025},
  organization={IEEE}
}

@article{nussbaum2025understanding,
  title={Understanding Voice Naturalness},
  author={Nussbaum, Christine and Fr{\"u}hholz, Sascha and Schweinberger, Stefan R.},
  journal={Trends in Cognitive Sciences},
  year={2025},
  publisher={Elsevier}
}

@inproceedings{dall2014rating,
  title={Rating Naturalness in Speech Synthesis: The Effect of Style and Expectation},
  author={Dall, Rasmus and Yamagishi, Junichi and King, Simon},
  booktitle={Speech Prosody 2014},
  year={2014}
}

@inproceedings{sellam2023squid,
  title={SQUID: Measuring Speech Naturalness in Many Languages},
  author={Sellam, Thibault and Bapna, Ankur and Camp, Joshua and Mackinnon, Diana and Parikh, Ankur P. and Riesa, Jason},
  booktitle={ICASSP 2023 - 2023 IEEE International Conference on Acoustics, Speech and Signal Processing (ICASSP)},
  pages={1--5},
  year={2023},
  organization={IEEE}
}

@article{vojtech2019effects,
  title={The Effects of Modulating Fundamental Frequency and Speech Rate on the Intelligibility, Communication Efficiency, and Perceived Naturalness of Synthetic Speech},
  author={Vojtech, Jennifer M. and Noordzij Jr, Jacob P. and Cler, Gabriel J. and Stepp, Cara E.},
  journal={American Journal of Speech-Language Pathology},
  volume={28},
  number={2S},
  pages={875--886},
  year={2019},
  publisher={American Speech-Language-Hearing Association}
}

@article{todisco2019asvspoof,
  title={ASVspoof 2019: Future Horizons in Spoofed and Fake Audio Detection},
  author={Todisco, Massimiliano and Wang, Xin and Vestman, Ville and Sahidullah, Md and Delgado, H{\'e}ctor and Nautsch, Andreas and Yamagishi, Junichi and Evans, Nicholas and Kinnunen, Tomi and Lee, Kong Aik},
  journal={arXiv preprint arXiv:1904.05441},
  year={2019}
}

@article{li2024audio,
  title={Audio Anti-Spoofing Detection: A Survey},
  author={Li, Menglu and Ahmadiadli, Yasaman and Zhang, Xiao-Ping},
  journal={arXiv preprint arXiv:2404.13914},
  year={2024}
}

@article{baser2025phonemefake,
  title={PhonemeFake: Redefining Deepfake Realism with Language-Driven Segmental Manipulation and Adaptive Bilevel Detection},
  author={Baser, Oguzhan and Tanriverdi, Ahmet Ege and Vishwanath, Sriram and Chinchali, Sandeep P.},
  journal={arXiv preprint arXiv:2506.22783},
  year={2025}
}

@inproceedings{temmar2025phonetic,
  title={Phonetic Analysis of Real and Synthetic Speech Using HuBERT Embeddings: Perspectives for Deepfake Detection},
  author={Temmar, Dia Elhak and Hamadene, Assia and Nallaguntla, Vamshi and Fursule, Aishwarya and Allili, Mohand Sa{\"\i}d and Kshirsagar, Shruti and Avila, Anderson R.},
  booktitle={2025 IEEE International Conference on Systems, Man, and Cybernetics (SMC)},
  pages={86--91},
  year={2025},
  organization={IEEE}
}

@article{yang2025forensic,
  title={Forensic Deepfake Audio Detection Using Segmental Speech Features},
  author={Yang, Tianle and Sun, Chengzhe and Lyu, Siwei and Rose, Phil},
  journal={Forensic Science International},
  pages={112768},
  year={2025},
  publisher={Elsevier}
}

@article{salvi2025phoneme,
  title={Phoneme-Level Analysis for Person-of-Interest Speech Deepfake Detection},
  author={Salvi, Davide and Negroni, Viola and Mandelli, Sara and Bestagini, Paolo and Tubaro, Stefano},
  journal={arXiv preprint arXiv:2507.08626},
  year={2025}
}

@inproceedings{jung2022aasist,
  title={AASIST: Audio Anti-Spoofing Using Integrated Spectro-Temporal Graph Attention Networks},
  author={Jung, Jee-weon and Heo, Hee-Soo and Tak, Hemlata and Shim, Hye-jin and Chung, Joon Son and Lee, Bong-Jin and Yu, Ha-Jin and Evans, Nicholas},
  booktitle={ICASSP 2022 - 2022 IEEE International Conference on Acoustics, Speech and Signal Processing (ICASSP)},
  pages={6367--6371},
  year={2022},
  organization={IEEE}
}

@inproceedings{zhang2025phoneme,
  title={Phoneme-Level Feature Discrepancies: A Key to Detecting Sophisticated Speech Deepfakes},
  author={Zhang, Kuiyuan and Hua, Zhongyun and Lan, Rushi and Zhang, Yushu and Guo, Yifang},
  booktitle={Proceedings of the AAAI Conference on Artificial Intelligence},
  pages={1066--1074},
  year={2025}
}

@article{Zhu2024SLIM,
  title={SLIM: Style-Linguistics Mismatch Model for Generalized Audio Deepfake Detection},
  author={Zhu, Yi and Koppisetti, Surya and Tran, Trang and Bharaj, Gaurav},
  journal={Advances in Neural Information Processing Systems},
  volume={37},
  pages={67901--67928},
  year={2024}
}

@inproceedings{McAuliffe2017MFA,
  title={Montreal Forced Aligner: Trainable Text-Speech Alignment Using Kaldi},
  author={McAuliffe, Michael and Socolof, Michaela and Mihuc, Sarah and Wagner, Michael and Sonderegger, Morgan},
  booktitle={Interspeech},
  pages={498--502},
  year={2017}
}

@article{chen2022wavlm,
  title={WavLM: Large-Scale Self-Supervised Pre-Training for Full Stack Speech Processing},
  author={Chen, Sanyuan and Wang, Chengyi and Chen, Zhengyang and Wu, Yu and Liu, Shujie and Chen, Zhuo and Li, Jinyu and Kanda, Naoyuki and Yoshioka, Takuya and Xiao, Xiong and others},
  journal={IEEE Journal of Selected Topics in Signal Processing},
  volume={16},
  number={6},
  pages={1505--1518},
  year={2022},
  publisher={IEEE}
}

@article{baevski2020wav2vec,
  title={wav2vec 2.0: A Framework for Self-Supervised Learning of Speech Representations},
  author={Baevski, Alexei and Zhou, Yuhao and Mohamed, Abdelrahman and Auli, Michael},
  journal={Advances in Neural Information Processing Systems},
  volume={33},
  pages={12449--12460},
  year={2020}
}

@book{hosmer2013applied,
  title={Applied Logistic Regression},
  author={Hosmer, David W. and Lemeshow, Stanley and Sturdivant, Rodney X.},
  year={2013},
  publisher={John Wiley \& Sons},
  edition={3rd},
  address={Hoboken, NJ}
}

@article{cortes1995support,
  title={Support-Vector Networks},
  author={Cortes, Corinna and Vapnik, Vladimir},
  journal={Machine Learning},
  volume={20},
  number={3},
  pages={273--297},
  year={1995},
  publisher={Springer}
}

@inproceedings{LibriSpeech,
  author={Panayotov, Vassil and Chen, Guoguo and Povey, Daniel and Khudanpur, Sanjeev},
  title={LibriSpeech: An ASR Corpus Based on Public Domain Audio Books},
  booktitle={2015 IEEE International Conference on Acoustics, Speech and Signal Processing (ICASSP)},
  pages={5206--5210},
  year={2015},
  organization={IEEE}
}

@misc{VCTK,
  author={Veaux, Christophe and Yamagishi, Junichi and MacDonald, Kirsten},
  title={CSTR VCTK Corpus: English Multi-Speaker Corpus for CSTR Voice Cloning Toolkit},
  year={2017},
  publisher={University of Edinburgh, Centre for Speech Technology Research (CSTR)},
  note={Available at: http://dx.doi.org/10.7488/ds/1994}
}

\end{document}